\PassOptionsToPackage{pdfpagelabels=false}{hyperref}
\documentclass[fleqn,usenatbib,usedcolumn]{mnras}
\usepackage[british]{babel}             

\usepackage{newtxmath,newtxtext}
\DeclareSymbolFont{operators}{OT1}{ntxtlf}{m}{n}
\SetSymbolFont{operators}{bold}{OT1}{ntxtlf}{b}{n}
\usepackage[T1]{fontenc}

\usepackage{graphicx}    
\usepackage{amsmath}    
\usepackage{xspace}
\usepackage{xstring}
\usepackage{pifont}
\usepackage{pgfkeys}

\newcommand{\gaia}{\textit{Gaia}\xspace}
\newcommand{\tess}{\textit{TESS}}
\newcommand{\Ktwo}{\textit{K2}}

\newcommand{\teff}{\ensuremath{T_\textrm{eff}}\xspace}
\newcommand{\logg}{\ensuremath{\log g}\xspace}
\newcommand{\feh}{\ensuremath{[\mathrm{Fe}/\mathrm{H}]}\xspace}
\newcommand{\alphafe}{\ensuremath{[\mathrm{\upalpha}/\mathrm{Fe}]}\xspace}

\newcommand{\alfe}{\ensuremath{[\mathrm{Al}/\mathrm{Fe}]}\xspace}
\newcommand{\mgfe}{\ensuremath{[\mathrm{Mg}/\mathrm{Fe}]}\xspace}
\newcommand{\kms}{\ensuremath{\textrm{km}\,\textrm{s}^{-1}}\xspace}

\newcommand{\ali}{\ensuremath{\mathrm{A}_\mathrm{Li}}\xspace}
\newcommand{\vbroad}{\ensuremath{v_\mathrm{broad}}\xspace}

\newcommand{\jphi}{\ensuremath{J_\phi}\xspace}
\newcommand{\jr}{\ensuremath{J_R}\xspace}
\newcommand{\jphiunits}{\ensuremath{\mathrm{Mpc}\,\mathrm{km}\,\mathrm{s}^{-1}}\xspace}
\newcommand{\jrunits}{\ensuremath{[\mathrm{kpc}\,\mathrm{km}\,\mathrm{s}^{-1}]^{1/2}}\xspace}
\newcommand{\nuclide}[2]{$^{#1}${#2}}

\title[Extragalactic Lithium Problem]{The GALAH Survey: Accreted stars also inhabit the Spite Plateau}

\author[J. D. Simpson et al.]{
\parbox{\textwidth}{\raggedright
Jeffrey~D.~Simpson,$^{1,2}$\thanks{Email: \texttt{jeffrey.simpson@unsw.edu.au}}
Sarah~L.~Martell,$^{1,2}$ 
Sven~Buder,$^{2,3}$ 
Joss~Bland\nobreakdash-Hawthorn,$^{2,4}$ 
Andrew~R.~Casey,$^{5,6}$ 
Gayandhi~M.~De~Silva,$^{7,8}$ 
Valentina~D'Orazi,$^{9}$ 
Ken~C.~Freeman,$^{2,3}$ 
Michael~Hayden,$^{2,4}$ 
Janez~Kos,$^{10}$ 
Geraint~F.~Lewis,$^{4}$ 
Karin~Lind,$^{11}$ 
Katharine~J.~Schlesinger,$^{3}$ 
Sanjib~Sharma,$^{2,4}$ 
Dennis~Stello,$^{1,2,4}$ 
Daniel~B.~Zucker,$^{2,8,12}$ 
Toma\v{z}~Zwitter,$^{10}$ 
Martin~Asplund,$^{13}$ 
Gary~Da~Costa,$^{2,3}$ 
Klemen~\v{C}otar,$^{10}$ 
Thor~Tepper\nobreakdash-García,$^{2,4,14}$ 
Jonathan~Horner,$^{15}$ 
Thomas~Nordlander,$^{2,3}$ 
Yuan\nobreakdash-Sen~Ting,$^{3,16,17,18}$ 
Rosemary~F.~G.~Wyse$^{19}$ 
and The~GALAH~Collaboration$^{}$ 
}
\\
\\
$^{1}$School of Physics, UNSW, Sydney, NSW 2052, Australia\\
$^{2}$Centre of Excellence for Astrophysics in Three Dimensions (ASTRO-3D), Australia\\
$^{3}$Research School of Astronomy \& Astrophysics, Australian National University, ACT 2611, Australia\\
$^{4}$Sydney Institute for Astronomy, School of Physics, A28, The University of Sydney, NSW, 2006, Australia\\
$^{5}$School of Physics and Astronomy, Monash University, Clayton, VIC 3800, Australia\\
$^{6}$Monash Centre for Astrophysics, School of Physics and Astronomy, Monash University, Australia\\
$^{7}$Australian Astronomical Optics, Faculty of Science and Engineering, Macquarie University, Macquarie Park, NSW 2113, Australia\\
$^{8}$Macquarie University Research Centre for Astronomy, Astrophysics \& Astrophotonics, Sydney, NSW 2109, Australia\\
$^{9}$Istituto Nazionale di Astrofisica, Osservatorio Astronomico di Padova, vicolo dell'Osservatorio 5, 35122, Padova, Italy\\
$^{10}$Faculty of Mathematics and Physics, University of Ljubljana, Jadranska 19, 1000 Ljubljana, Slovenia\\
$^{11}$Department of Astronomy, Stockholm University, AlbaNova University Centre, SE-106 91 Stockholm, Sweden\\
$^{12}$Department of Physics and Astronomy, Macquarie University, Sydney, NSW 2109, Australia\\
$^{13}$Max Planck Institute for Astrophysics, Karl-Schwarzschild-Str. 1, D-85741 Garching, Germany\\
$^{14}$Centre for Integrated Sustainability Analysis, The University of Sydney\\
$^{15}$Centre for Astrophysics, University of Southern Queensland, Toowoomba, QLD 4350, Australia\\
$^{16}$Institute for Advanced Study, Princeton, NJ 08540, USA\\
$^{17}$Department of Astrophysical Sciences, Princeton University, Princeton, NJ 08544, USA\\
$^{18}$Observatories of the Carnegie Institution of Washington, 813 Santa Barbara Street, Pasadena, CA 91101, USA\\
$^{19}$Department of Physics and Astronomy, Johns Hopkins University, Baltimore, MD 21218, USA\\
}

\date{Accepted XXX. Received YYY; in original form \today}

\pubyear{2021}

\begin{document}
\label{firstpage}
\pagerange{\pageref{firstpage}--\pageref{lastpage}}
\maketitle

\begin{abstract}
The ESA \gaia mission has enabled the remarkable discovery that a large fraction of the stars near the Solar neighbourhood are debris from a single in-falling system, the so-called \gaia-Enceladus-Sausage (GSE). This discovery provides astronomers for the first time with a large cohort of easily observable, unevolved stars that formed in a single extra-Galactic environment.
Here we use these stars to investigate the ``Spite Plateau'' --- the near-constant lithium abundance observed in unevolved metal-poor stars across a wide range of metallicities ($-3<\feh<-1$). Our aim is to test whether individual galaxies could have different Spite Plateaus --- e.g., the interstellar medium could be more depleted in lithium in a lower galactic mass system due to it having a smaller reservoir of gas.
We identified 93 GSE dwarf stars observed and analyzed by the GALactic Archeology with HERMES (GALAH) survey as part of its Third Data Release. Orbital actions were used to select samples of GSE stars, and comparison samples of halo and disk stars.
We find that the GSE stars show the same lithium abundance as other likely accreted stars and \textit{in situ} Milky Way stars. Formation environment leaves no imprint on lithium abundances. 
This result fits within the growing consensus that the Spite Plateau, and more generally the "cosmological lithium problem" --- the observed discrepancy between the amount of lithium in warm, metal-poor dwarf stars in our Galaxy, and the amount of lithium predicted to have been produced by Big Bang Nucleosynthesis --- is the result of lithium depletion processes within stars.
\end{abstract}



\section{Introduction}\label{sec:intro}

\nuclide{7}{Li} is one of the four stable nuclides (along with \nuclide{2}{H}, \nuclide{3}{He}, \nuclide{4}{He}) that are predicted by Big Bang Nucleosynthesis (BBN) to have been synthesized during the first minutes of the Universe \citep[a theory first proposed by][see recent reviews of BBN from \citealt{Cyburt2016} and \citealt{Fields2020}]{Alpher1948}. In the standard model of BBN, the amounts of these primordial abundances formed are strongly dependent on the baryon-to-photon ratio, which can be estimated from the Cosmic Microwave Background. Observations in astrophysical environments have measured abundances that are in agreement with BBN predictions \citep{Fields2020} i.e., {\nuclide{2}{H} from quasar spectra \citep[e.g.,][]{Cooke2018}, \nuclide{4}{He} from metal-poor extra-galactic HII regions \citep[e.g.,][]{Aver2015}, \nuclide{7}{Li} in low metallicity gas in the Small Magellanic Cloud \citep{Howk2012} --- though the latter is dependent on the chemical evolution model.}

There has long been tension between the BBN-predicted primordial abundance of \nuclide{7}{Li} and the abundance measured in warm, metal-poor dwarf stars of our Galaxy. These stars have thin convective layers, and models predict they formed from gas that had a (near-)primordial Li abundance, and they will retain this formation Li abundance in their photosphere \citep{Deliyannis1990}. As first identified by \citet[][\citeyear{Spite1982a}]{Spite1982}, and subsequently confirmed \cite[e.g.,][]{Bonifacio1997,Ryan1999,Asplund2006,Sbordone2010,Melendez2010}, across a large range of metallicity ($-3<\feh<-1$) these stars do have a near-constant Li abundance creating the so-called ``Spite Plateau''. But, the Spite Plateau abundance is 3--4 times lower than \nuclide{7}{Li} abundance predicted by BBN. This discrepancy is known as the ``cosmological lithium problem'' \citep[see the review by][]{Fields2011}.

The concordance of the predictions from BBN and the observed abundances {for the \nuclide{2}{H}, \nuclide{3}{He}, and \nuclide{4}{He} nuclides} means that it is not BBN at fault. There is a growing consensus that the Li depletion in these stars is from extra mixing below the convection zone due to rotation \citep[e.g.,][]{Deliyannis1990,Pinsonneault1992}, diffusion and turbulent mixing \citep[e.g.,][]{Michaud1991,Richard2012}, and internal gravity waves \citep[e.g.,][]{Charbonnel2005}.

In this work, we investigate an alternative hypothesis to explain the Spite Plateau: that observable metal-poor dwarf stars simply did not form with a near-primordial lithium abundance. The Spite Plateau could be the result of pre-processing of interstellar medium through previous generations of stars, as proposed by, e.g., \citet{Piau2006}. This raises the question that we wish to answer: \textit{does the stellar formation environment affect the Spite Plateau?}

Unfortunately, it is not practical to measure the lithium abundance of dwarf stars in external galaxies with the current suite of telescopes\footnote{In the future it will be possible, e.g., G-CLEF on the Giant Magellan Telescope will be able to acquire in 1.5~hours $R\approx23000$ spectra of $\mathrm{SNR}{\sim}10$ for the brightest turn-off stars in the LMC.}. Fortunately, it is thought that a sizeable fraction of the Galactic halo stars were accreted \citep[e.g.,][and reviews by \citealt{Bland-Hawthorn2016} and \citealt{Helmi2020}]{Ibata1994, Helmi1999, Ibata2004, Abadi2010, Starkenburg2017}, with the accepted paradigm for the growth of galaxies like our Milky Way featuring the hierarchical mergers of large galaxies with smaller dwarf galaxies \citep[e.g.,][]{Searle1978,White1991,Cole2002,Venn2004,Oser2010}. These dwarf galaxies are completely disrupted and are mixed throughout the Galactic halo. 

Until recently, globular clusters provided the only dwarf stars that could be easily identifiable as being born \textit{ex situ} that are observationally-accessible. For instance, dwarf stars in $\upomega$~Cen show lithium abundances consistent with field stars \citep{Monaco2018}. Using models of diffusion to estimate their initial abundances, \citet{Mucciarelli2014} observed giant stars in M54, finding the Galactic Spite Plateau value. We do note two complications with globular clusters: firstly, the question of which globulars are accreted and which formed \textit{in situ} is an active area of research \citep[e.g.,][]{Forbes2010,Massari2019,Horta2020,Keller2020}; and secondly, the observations that the first and second generations of stars have the same lithium abundances, despite the latter stars forming from material processed through the former \citep[see discussion in the review by][]{Gratton2019}.

Now, thanks to the combination of deep photometric surveys and the data contained in the Second Data Release of the ESA \gaia mission \citep{GaiaCollaboration2016,GaiaCollaboration2018b}, it has been shown that the halo of the Galaxy contains spatial and kinematic substructure --- the identifiable remnants of past accretion events \citep[][and see the review by \citealt{Helmi2020}]{Malhan2018,Ibata2018,Yan2018,Barba2019,Myeong2019,Yuan2020,Naidu2020,Borsato2020}. Some stars of the $>50$ known ``stellar streams'' \citep[][and references therein]{Mateu2018} have been spectroscopically observed \citep[e.g.,][]{Li2019,Ibata2019a,Simpson2020,Li2020a,Ji2020}. For the S2 stream, the likely result of a disrupted dwarf galaxy, \citet{Aguado2021} measured the lithium abundance for a handful of dwarf stars and found the lithium behaviour to be like that of the rest of the halo.

On a larger scale, a significant proportion of the halo stars near the Sun appear to have been accreted from a single dwarf galaxy \citep[e.g.,][]{Nissen2010,Helmi2018,Myeong2018a,Myeong2018b,Haywood2018,Fattahi2019,Belokurov2018,Naidu2020}. ``Gaia-Enceladus-Sausage'' (GSE) presents the exciting prospect of a large number of main-sequence stars from a single dwarf galaxy that can be readily observed with current ground-based telescopes. These stars have a relatively unique kinematic signature --- eccentric orbits with low angular momenta --- making them easy to distinguish from other halo stars. For moderately metal-poor members ($\feh\geq-1.3$) they can also be chemically tagged as coming from the GSE due to their low \alphafe compared to \textit{in situ} stars \citep[][]{Nissen2010,Helmi2018,Monty2020}.

Previous studies have looked at lithium in the GSE stars. \citet{Nissen2012} measured the lithium abundances of 25 stars in their low-$\upalpha$ halo sequence --- which we now know to be GSE --- finding no significant systematic difference in the lithium abundances of high- and low-$\upalpha$ stars. \citet{Molaro2020} and \citet{Cescutti2020} both looked at lithium in 39 GSE stars using abundances from literature compilations. They found that the Spite Plateau in GSE to be the same as in the rest of the Milky Way.

In this work we expand upon those results, presenting the lithium abundance of 93 GSE dwarf stars serendipitously observed and homogeneously analyzed as part of the Third Data Release \citep{Buder2021} of the GALactic Archeology with HERMES survey \citep[GALAH;][]{DeSilva2015}. GALAH is a massive spectroscopic survey of the local Galactic volume, and we explore whether the lithium abundances of GSE stars are consistent with the abundances of other Galactic populations with likely \textit{in situ} and \textit{ex situ} origins. We test the hypothesis that the Spite Plateau is not the result of galactic chemical evolution.

This work is structured as follows: Section \ref{sec:observations} describes the observation, reduction, and analysis; Section \ref{sec:lithium} considers how lithium abundances change with stellar properties; Section \ref{sec:ges_selection} explains how the GSE stars were selected; Section \ref{sec:ges_spite} compares and contrasts the Spite Plateau for various sub-samples of Milky Way stars; Section \ref{sec:enrichment} comments on a handful of very Li-rich metal-poor stars; Section \ref{sec:discussion} summarizes the paper.

\section{Observation, reduction, and analysis}\label{sec:observations}
The spectroscopic data used in this work comes from the 588,571 stars of the Third Data Release (DR3) of the GALAH survey \citep{Buder2021} --- the combination of GALAH survey \citep{DeSilva2015,Martell2017,Buder2018}, the \Ktwo-HERMES survey \citep{Wittenmyer2018,Sharma2019} and the \tess-HERMES survey \citep{Sharma2018}. All observations used the HERMES spectrograph \citep{Sheinis2015} and the 2dF fibre positioning system \citep{Lewis2002} at the 3.9-m Anglo-Australian Telescope. HERMES records ${\sim}1000$~\AA\ of the optical spectrum at a spectral resolution of $R \approx 28000$ across four non-contiguous sections, which includes the neutral Li resonance lines at 6708~\AA. The spectra were reduced with a custom \textsc{iraf} pipeline \citep{Kos2017}.

\begin{figure}
	\includegraphics[width=\columnwidth]{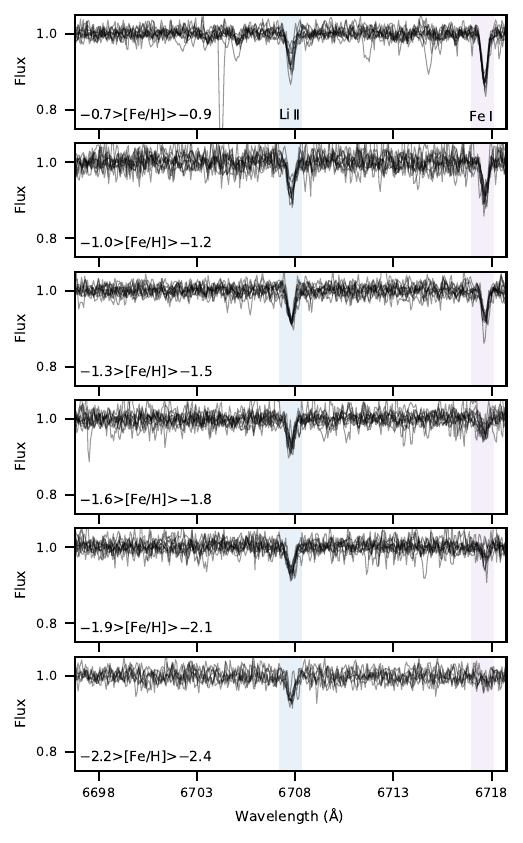}
    \caption{Pseudo-continuum normalized HERMES spectra for the Li~6708~\AA\ region (shaded blue band). In each panel there are up to 10 randomly chosen stars in a narrow \feh range (as indicated on each panel) from a representative region of the \teff-\logg space for dwarf stars ($5975~\mathrm{K}<\teff<6175~\mathrm{K}$ and $4.0 < \logg < 4.4$). The purple shaded band is the Fe~6718~\AA\ line seen to weaken with decreasing metallicity.  There is little variation in the strength of the lithium line in a given metallicity bin, qualitatively confirming the presence of the Spite Plateau. \label{fig:comparison_spec_plots}}
\end{figure}

{All stellar parameter and abundance values in this work are from \citet{Buder2021}, with the} lithium abundance for each star determined from synthesis of the 6708~\AA\ Li line {and includes non-LTE corrections}. In this work we use the form $\ali=\log[n_\mathrm{Li}/n_\mathrm{H}]+12$, where $n_\mathrm{Li}$ and $n_\mathrm{H}$ are the number densities of lithium and hydrogen respectively. On this scale, the BBN prediction is $\ali=2.75\pm0.02$ \citep{Pitrou2018}. Examples of HERMES spectra of dwarf stars for the region around 6708~\AA\ are shown in Figure~\ref{fig:comparison_spec_plots}. Qualitatively, these spectra confirm the existence of the Spite Plateau --- i.e., at a given \feh and \teff there is little variation in the strength of the lithium line. 

We apply data quality selections to identify a sample of dwarf stars (defined as surface gravity $\logg>3.65$ and absolute G magnitude~$>1.5$) with reliable stellar parameters and abundances. For each star we require: (i) GALAH DR3 flag $\texttt{flag\_sp}=0$ and $\texttt{flag\_fe\_h}=0$ (no problems noted in the input data, reduction, analysis, or iron abundance determination); (ii) a five-parameter solution from \gaia eDR3 \citep{GaiaCollaboration2021} to allow for orbital calculations; (iii) the red camera spectrum (which contains the Li line) signal-to-noise ratio $>30$ per pixel. These selections identify the set of 197,921 dwarf stars that we consider for the remainder of this work. When we are considering the abundance of element $\texttt{x}$, we also require that the GALAH flag $\texttt{flag\_x\_fe}=0$ (no problems noted in the abundance determination). For instance, of the 197,921 dwarf stars, 86,320 dwarf stars with reliable lithium abundances.

The Galactic orbital properties of each star are taken from \citet{Buder2021}. Briefly, this used \textsc{galpy} {\citep{Bovy2015}} with values from GALAH DR3 and \gaia eDR3, \textsc{McMillan2017} potential \citep{McMillan2017} and the values $R_\mathrm{GC}=8.21$~kpc and $v_\mathrm{circular}=233.1~\kms$ \citep{TheGRAVITYCollaboration2019}. It set $(U,V,W)_\mathrm{\odot}=(11.1, 15.17, 7.25)~\kms$ in keeping with \citet{Reid2004} and \citet{Schonrich2010}. Distances are primarily from the GALAH DR3 age and mass value-added catalogue, which mostly incorporated distances found by the Bayesian Stellar Parameters estimator \citep[\textsc{bstep}; described in][]{Sharma2018}.

\section{Effects of stellar properties on lithium abundances}\label{sec:lithium}

\begin{figure}
	\includegraphics[width=\columnwidth]{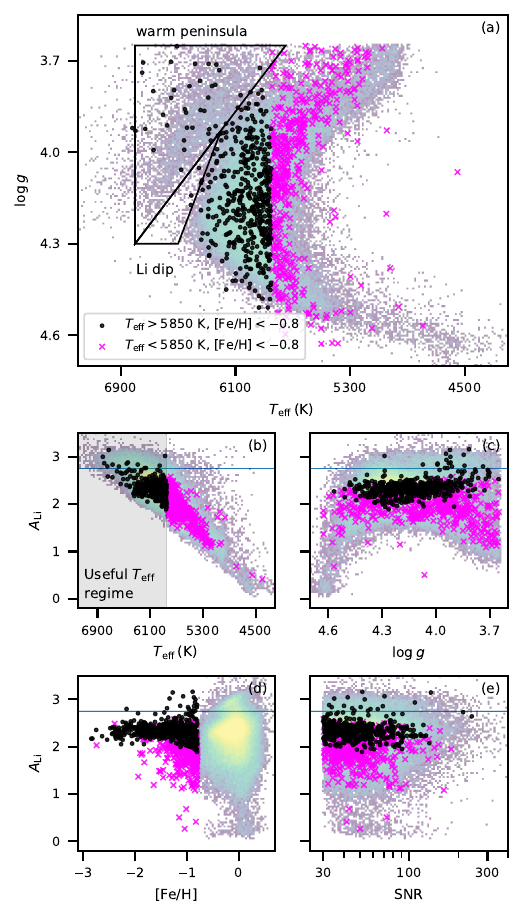}
    \caption{The distribution of the 86,320 GALAH stars with reliable lithium abundances (i.e., $\texttt{flag\_li\_fe}=0$) in (a) \teff-\logg; and \ali versus (b) \teff, (c) \logg, (d) \feh, and (e) SNR. The horizontal blue line is the BBN lithium abundance \citep[$\ali=2.75\pm0.02$;][]{Pitrou2018}. Highlighted are all stars with $\feh<-0.8$, with black dots for stars with $\teff>5850$~K, and fuchsia crosses for stars with $\teff<5850$~K. The hotter stars are our selection of metal-poor stars that are least affected by lithium depletion on the main sequence. The cooler stars might otherwise have been thought to form part of the Spite Plateau, but can be seen to have much lower lithium abundances. As this Figure only shows stars with reliable lithium abundances, the morphology of the \teff-\logg plane shows two interesting features, highlighted with triangular regions on (a): the ``warm peninsula'' of Li-rich dwarf stars, and the ``Li dip'' where there is a dearth of stars with lithium measurements.\label{fig:lithium_selection}}
\end{figure}

Lithium is a fragile element in stars. Along with Be and B \citep[see the review of][]{Randich2021} it is easily destroyed in stellar interiors at a relatively low temperature \citep[$2.6\times 10^6$~K for Li;][]{Gamow1933,Salpeter1955}. This destruction is observed in the Sun: the Solar photospheric lithium abundance is $\ali=1.05\pm0.01$, while the meteoritic lithium abundance is $3.26\pm0.05$ \citep{Asplund2009}. Li burning will occur in stars whenever there is the ability to transport surface material to the hotter interior. There are various stellar atmospheric processes --- the convective zone, atomic diffusion \citep[e.g.,][]{Michaud1991,Richard2012}, rotation-induced mixing \citep[e.g.,][]{Deliyannis1990,Pinsonneault1992}, and internal gravity waves \citep[e.g.,][]{Charbonnel2005} --- competing to both drive and inhibit this mixing. The dominant mechanism depends on the mass, metallicity, and evolutionary stage of the star. For this work it is necessary to identify stars that are the least affected by Li depletion, and therefore their current Li abundance is most representative of their birth abundance.

Figure~\ref{fig:lithium_selection} shows the 86,320 dwarf stars from our sample with reliable lithium abundance (e.g., $\mathtt{flag\_li_\_fe}=0$) in the \teff-\logg plane, and their \ali with respect to \teff, \logg, \feh, and signal-to-noise in the HERMES spectra. The coolest stars, with the highest and lowest \logg in the sample, likely experienced a large amount of surface lithium depletion and form ``tails'' of low \ali in Figure~\ref{fig:lithium_selection}c. The hottest stars retain most of their formation lithium because they have thin convective zones and are nearly unaffected by either rotation-induced mixing or diffusion. Seen in Figure~\ref{fig:lithium_selection}a as the ``warm peninsula'', these stars also form the over-density of roughly solar metallicity stars with $\ali\sim3$ in Figure~\ref{fig:lithium_selection}d. Meanwhile for stars in the range $6400~\mathrm{K}<\teff<6850$~K, rotation-induced mixing is the dominant mechanism, causing severe lithium destruction and the so-called ``lithium dip'' \citep[a feature first observed in open clusters, e.g.,][]{Wallerstein1965,Boesgaard2016}. This Li-dip manifests as the region of the \teff-\logg diagram lacking in almost any stars with reliable lithium measurements. The Li-dip and ``warm peninsula'' are explored in detail using GALAH data by \citet{Gao2020}.

For investigating the Spite Plateau, as has been done previously in the literature \citep[e.g.,][]{Melendez2010}, we want to identify a stellar parameter selection that includes only those stars hot enough to inhibit significant Li depletion, whilst simultaneously maximizing our sample size of metal-poor stars. Experimentation identified a temperature cut-off of $\teff>5850$~K as the best compromise between selecting those stars least affected by Li depletion, and retaining a large sample of stars. In Figure~\ref{fig:lithium_selection}, we highlight stars with $\feh<-0.8$ (i.e., the metallicity range of halo stars): black dots for stars with $\teff>5850$~K, and fuchsia crosses for stars below this temperature. There are a handful of hotter metal-poor stars that fall onto the Li-rich ``warm peninsula'' --- the stars least affected by any Li depletion. However, as we will see in Section~\ref{sec:ges_spite}, none of these stars are found in our halo samples. Of the 86,320 dwarf stars with reliable Li abundances in our dataset, 53,761 have $\teff>5850$~K. For the 992 dwarf stars with metallicity $\feh<-0.8$, 485 have $\teff>5850$~K. 

\section{Gaia-Enceladus-Sausage member selection}\label{sec:ges_selection}
One of the major discoveries facilitated by \gaia is that about half of the metal-poor stars in the local halo appear to have been accreted from single dwarf galaxy called ``GSE'' \citep{Helmi2018,Belokurov2018,Myeong2018a,Haywood2018}. Here, we identify likely accreted GSE stars using the same method as \citet{Feuillet2020}, who cleanly selected GSE members as those with Galactic orbits that had angular momentum $\jphi\sim0$ and large radial action $\jr$. 

\begin{figure}
	\includegraphics[width=\columnwidth]{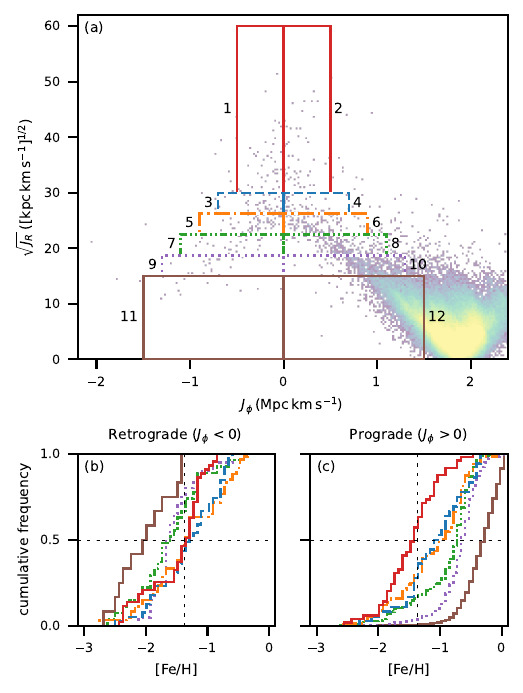}
    \caption{The cumulative metallicity distribution functions (CDF) for bins of stars in \jphi-$\sqrt{\jr}$ action space. (a) shows the log-density number distribution of all GALAH DR3 dwarf stars in \jphi-$\sqrt{\jr}$ action space, along with the location of the bins used for each CDF. (b) and (c) show respectively the CDF for the retrograde (odd-numbered bins; $\jphi<0$) and prograde (even-numbered bins; $\jphi>0$). The vertical dashed lines on (b) and (c) mark the median metallicity of the combination of all stars in bins 1 and 2 ($\feh=-1.37$). The horizontal line is for the 50th percentile of the CDF. The CDF for bins 1 and 2 (red on all panels) are the most consistent of all the pairings of prograde and retrograde bins; these bins form our selection of GSE stars.
     \label{fig:Feuillet_reproduction}}
\end{figure}

Figure~\ref{fig:Feuillet_reproduction} is similar to figure 5 from \citet{Feuillet2020}, but for our sample of 197,921 GALAH DR3 dwarfs. The \jphi-$\sqrt{\jr}$ space is divided into pairs of prograde ($\jphi>0$) and retrograde ($\jphi<0$) bins for a range of $\sqrt{\jr}$. As \citet{Feuillet2020} noted, stars with $\sqrt{\jr}>30~\jrunits$ (i.e., our bins 1 and 2, which will form our GSE selection) have similar cumulative metallicity density function (CDF) for both the prograde and retrograde bins --- consistent with them having a single origin. Meanwhile, in all bins other than 1 and 2, ($\sqrt{\jr}<30~\jrunits$) each pairing of the prograde and retrograde bins show different CDFs. All of the prograde bins are more metal rich than bins 1 and 2, due to the formers' inclusion of more disk-like (and therefore more metal rich) stars. The combined sample of stars from bins 1 and 2 have a median metallicity of $\feh=-1.37$, which is marked with dashed vertical lines on Figure~\ref{fig:Feuillet_reproduction}b and c. This is more metal poor than the median metallicity of $\feh=-1.17$ found by \citet{Feuillet2020} for the same kinematic selection, but their sample was giants with photometric metallicity estimates.

\begin{figure*}
	\includegraphics[width=\textwidth]{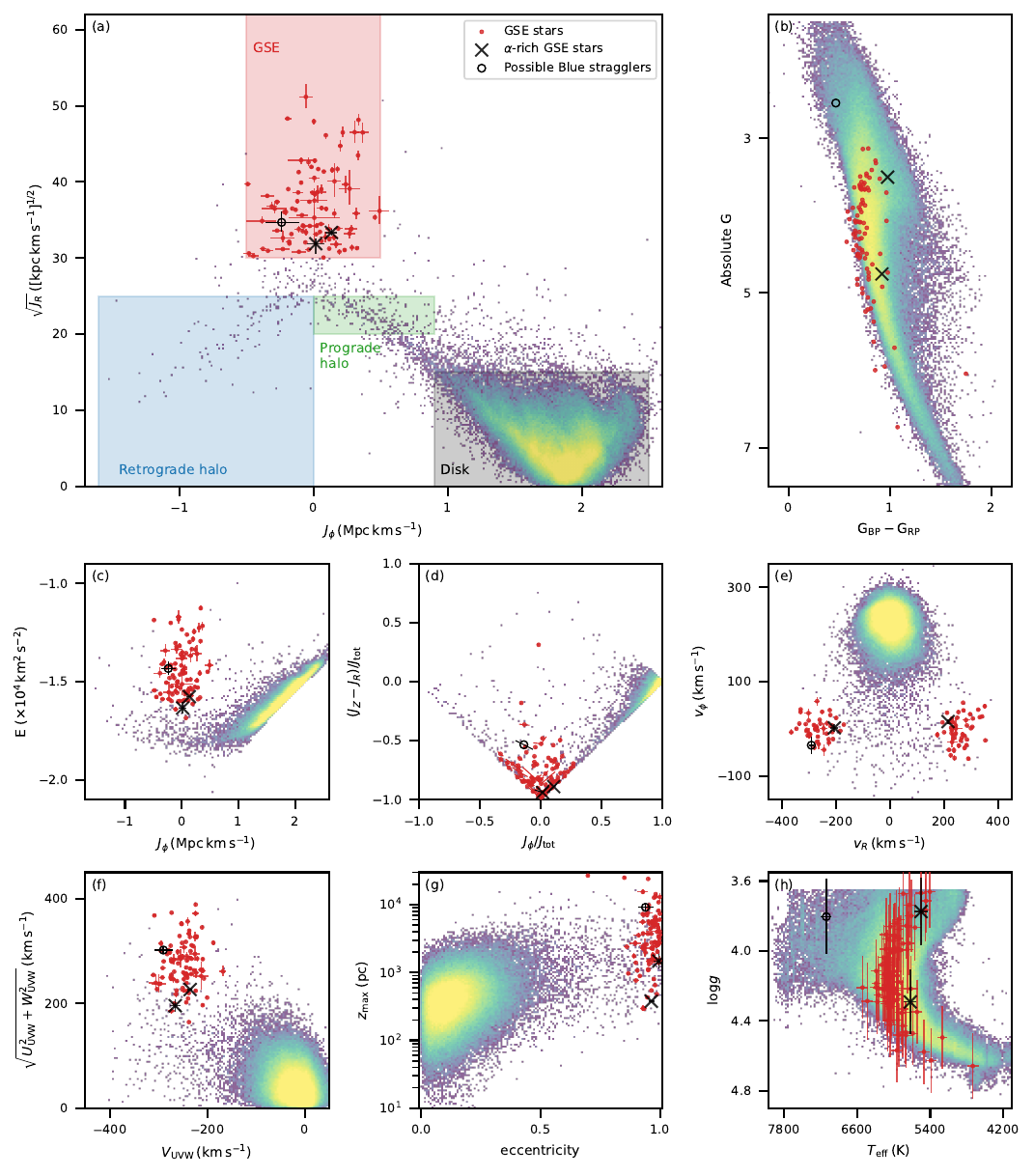}
    \caption{Distributions of all GALAH dwarfs in a variety of dynamical, orbital, and stellar parameter spaces that have been used to identify GSE stars: (a) the \jphi-$\sqrt{\jr}$ action-angle space, (b) the colour-(absolute) magnitude diagram, (c) \jphi and orbital energy, (d) \jphi and ($J_Z-\jr$) normalized by $J_\mathrm{tot}(=J_Z+\jr+|\jphi|)$, (e) the {Galactocentric velocity components} $v_r$ and $v_\phi$, (f) the Toomre diagram, (g) orbital eccentricity and the largest distance travelled out of the Galactic plane $z_\mathrm{max}$, and (h) \teff-\logg. In panel (a) the selections used to identify the various dynamical sub-samples are shown, and the stars within the GSE box are highlighted in red in all panels, except for three GSE stars that are highlighted in black. These latter stars are either much hotter than all other stars (possibly blue straggler stars; black open circle symbols); or are $\upalpha$-rich stars (see Figure~\ref{fig:abundance_comparison}; black crosses). \label{fig:initial_ges_selection_dynamics}}
\end{figure*}

We combine bins 1 and 2 to be the selection of GSE stars and highlight those stars in a variety of observational spaces on Figure~\ref{fig:initial_ges_selection_dynamics}. Three additional groups of stars are identified for comparison to our GSE sample, and these regions are indicated on Figure~\ref{fig:initial_ges_selection_dynamics}a. The details of these selections are as follows:
\begin{itemize}
    \item 93 GSE stars: $(-0.5<\jphi<0.5)~\jphiunits$ and $\sqrt{\jr}>30~\jrunits$;
	\item 102 retrograde orbiting halo stars: $\jphi<0~\jphiunits$ and $\sqrt{\jr}<25~\jrunits$;
	\item 208 prograde orbiting stars that will be a mixture of halo and dynamically thick disk stars: $[0<\jphi<1]~\jphiunits$ and $20<\sqrt{\jr}<25~\jrunits$;
	\item 196,619 disk stars, which are highly likely to have formed  \textit{in situ}: $\jphi>0.9~\jphiunits$ and $\sqrt{\jr}<15~\jrunits$.
\end{itemize}

{Different authors have used different selections for identifying members of GSE. The \jr-\jphi selection used in this work has much overlap with these other selections, but has the advantage of having less contamination from non-GSE stars.} Here we comment on the similarities and differences between the various selections used in other works:
\begin{itemize}
    \item \citet{Belokurov2018} used the $v_R$ versus $v_\phi$ plane to identify their ``Sausage'' of stars at $v_\phi\sim0$~\kms. Their selection includes stars at all $v_R$, rather than our selection which is limited to $|v_R|>\sim200$~\kms (Figure~\ref{fig:initial_ges_selection_dynamics}e). Having a lower limit in \jr for GSE would result in stars with smaller $|v_R|$ being included, while maintaining the range of $v_\phi$.
    \item \citet{Koppelman2018} and \citet{Helmi2018} identified their GSE stars as a ``plume'' of stars in the $\jphi$-Energy space (Figure~\ref{fig:initial_ges_selection_dynamics}c) and in the Toomre diagram (Figure~\ref{fig:initial_ges_selection_dynamics}f), which is where our GSE selection stars can also be found. The region used by \citet{Helmi2018} was more retrograde and includes stars that are found on the ``arm'' of stars that extend to low $\sqrt{\jr}$ and negative \jphi region of Figure~\ref{fig:initial_ges_selection_dynamics}a that we have defined as being the ``retrograde halo''.
    \item \citet{Myeong2019} selected their GSE sample using the action space map (Figure~\ref{fig:initial_ges_selection_dynamics}d). Their selection translates to a narrower range of angular momentum ($|\jphi|\leq0.1~\jphiunits$), but like \citet{Belokurov2018}, to a lower value of \jr than we have used.
    \item Our selection of stars has identified stars along a narrow sequence in the colour-magnitude diagram (Figure~\ref{fig:initial_ges_selection_dynamics}b) which is consistent with the blue sequence found by \citet{Haywood2018} and others.
\end{itemize}  

\subsection{Possible contamination of the sample}

\begin{figure}
	\includegraphics[width=\columnwidth]{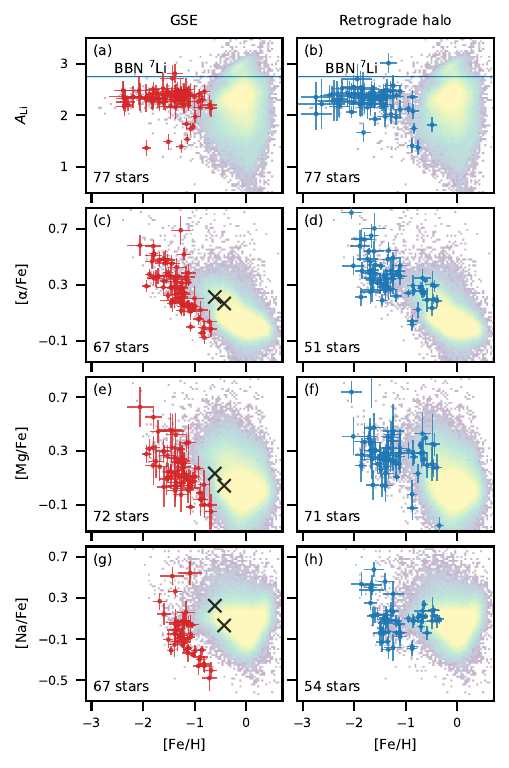}
    \caption{Abundance distributions for \ali, \alphafe, \mgfe, and [Na/Fe] for the GSE stars (left column; red dots with error bars) and retrograde halo sample (right column; blue dots with error bars), compared to the full sample of dwarf stars used in this work (background log-density distribution). The black symbols are the same as Figure~\ref{fig:initial_ges_selection_dynamics}, indicating dynamically selected GSE stars that are either $\upalpha$-rich (crosses) or much hotter than all other GSE stars (unfilled circles). The blue horizontal lines in panels (a) and (b) indicate the BBN prediction for \ali \citep[$\ali=2.75\pm0.02$;][]{Pitrou2018}. For the other elements, comparing the GSE sample to the retrograde halo sample, for stars with $\feh>\sim-1$, the GSE stars are generally all lower in abundance, while the retrograde halo stars have abundances more consistent with the $\upalpha$-rich disk.
     \label{fig:abundance_comparison}}
\end{figure}

In this subsection we consider if there is any contamination of our selection of GSE stars with non-member stars.

We have not used abundance information as a chemical tag of GSE stars \citep[e.g., Mg, Mn, and Al as used by][]{Das2020}, as this would limit us to only the metal-rich stars, due to the limitations of HERMES spectra for metal-poor stars. In the metallicity range $-2.5<\feh<-2.0$, there are 95 dwarf stars in GALAH DR3 with reliable parameters, of which 59 (62 per cent) have a reliable \ali, but only 16 (17 per cent) have \alphafe, 10 (11 per cent) have [Mg/Fe], 10 (11 per cent) have [Mn/Fe], and none have \alfe.

It is still useful to consider the abundance for the metal-rich end of the GSE selection. Figure~\ref{fig:abundance_comparison} shows a comparison of the abundance distributions of \ali, \alphafe, \mgfe, and [Na/Fe] 
for GSE (left column) and the stars from the retrograde halo sample (right column). $\upalpha$-elements like Mg 
and the odd-Z elements like Na have well-known ``knee'' in the Tinsley-Wallerstein diagram \citep{Buder2021} when SN Ia began to dominate SN II and decrease the abundance of some light elements in the interstellar medium \cite[e.g.,][]{Kobayashi2020}. For the Milky Way this is at $\feh\sim-1$, while lower galactic mass systems have their knee at a lower metallicity \citep{Venn2004}. We confirm this previously observed feature of GSE \citep{Helmi2018,Monty2020}. We find only two GSE stars (marked with black crosses on Figures \ref{fig:initial_ges_selection_dynamics} and \ref{fig:abundance_comparison}) that have \alphafe values clearly more consistent with the $\upalpha$-rich disk.

There is one further possible contaminant star. This star, marked with open black circle on Figure \ref{fig:initial_ges_selection_dynamics} has a location on \teff-\logg diagrams and CMD (Figures \ref{fig:initial_ges_selection_dynamics}b and h respectively) that makes the star too young and/or massive to belong to GSE. Briefly, there are three possible explanations: (1) it has relatively large errors in its orbital parameters, so it could simply be the result of unreliable measurements; (2) it is a young in situ Milky Way star that has ended up in a halo orbits by some dynamical process \citep{Famaey2005,Bensby2007,Williams2011,Casey2014}; (3) it is a blue straggler (BS) star. BS are main sequence-like stars that stand out in roughly-coeval populations because they are significantly more massive than the normal MS turn-off mass of the population \citep{Bailyn1995}. Many globular clusters have a population of BS, and the leading mechanisms invoked to explain their presence are binary mergers and/or stellar collisions. Identifying them in the field is much more difficult because you are no longer considering a `closed box' environment, but estimates suggest the fraction of the nearby halo BS is of the order of 20 per cent \citep{Casagrande2020}.

Overall, there appears to be little obvious contamination of non-GSE stars into our sample. Neither the possible blue stragglers, nor the $\upalpha$-rich stars affect the analysis of the Spite Plateau as they, by chance, lack reliable lithium abundances.

\section{The lithium plateau and enrichment in GSE}\label{sec:ges_spite}

\begin{table*}
 \caption{{Stellar parameters and lithium abundances for those stars identified as belonging the Spite Plateau in GALAH DR3. Here we show only the four stars from the GSE and retrograde halo samples that appear to be lithium enrichment (Figure~\ref{fig:possible_enrichment}). The full table is available as supplementary material.}}
 \label{table:the_stars}
 \begin{tabular}{rrrrrrrr}
  \hline
  Star ID & sobject\_id & dr3\_source\_id & \teff & \logg & \feh & \ali & Population \\
  \hline
08295313+1621105 & 160110002601062 & 658743190301110144 & 5988 & 3.98 & -1.43 & 2.72 & GES\\
13433203-3940090 & 190223003301120 & 6113351720646486528 & 5956 & 3.84 & -1.38 & 2.81 & GES\\
03355522-6833454 & 131116000501386 & 4667364088963367808 & 5874 & 4.29 & -1.33 & 3.02 & Retrograde halo\\
15123917-1944545 & 170507008301377 & 6256414985329648384 & 6338 & 4.29 & -1.94 & 2.70 & Retrograde halo\\
  \hline
 \end{tabular}
\end{table*}

\begin{figure*}
	\includegraphics[width=\textwidth]{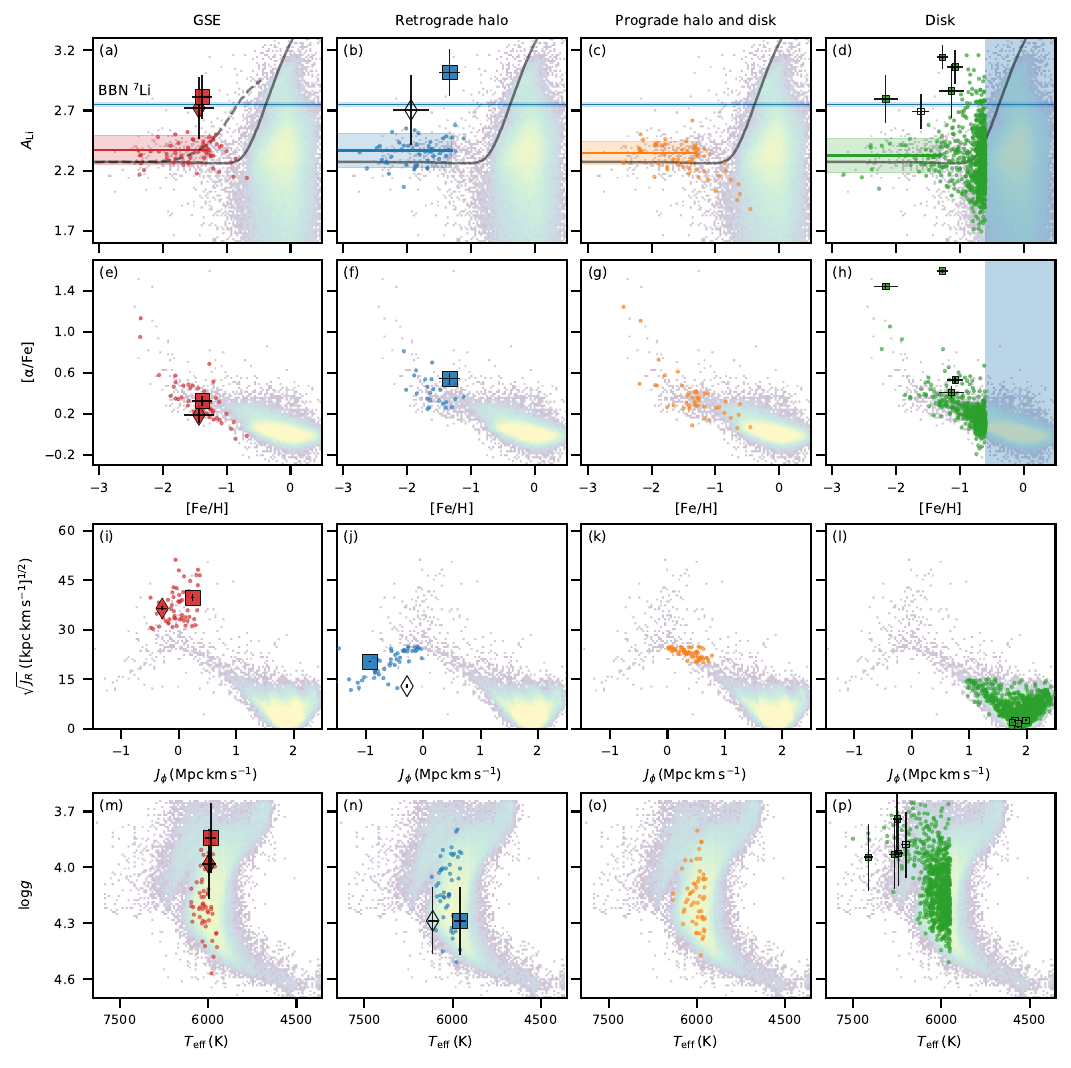}
    \caption{The (top row) \ali-\feh, (second row) \alphafe-\feh, \jphi-\jr (third row), and (bottom row) \teff-\logg distributions for the four dynamically selected populations of stars --- GSE (first column), retrograde halo (second column), prograde halo (third column), and disk (fourth column). The background distribution on all panels is all dwarf stars in GALAH with reliable lithium abundances. For each group the stars that meet the \teff-age criteria (Section \ref{sec:lithium} and Figure~\ref{fig:lithium_selection}) are highlighted. For visual clarity, the disk sample has been truncated to only show stars with $\feh<-0.6$. In the top row, the coloured horizontal line and shaded region indicate the average and standard deviation of stars with $\feh<-1.3$ --- i.e., the Spite Plateau for that population. Highlighted with black-edged squares or diamond symbols (filled for stars with a reliable \alphafe; different symbols on the GSE and retrograde halo is simply to help the reader distinguish the two stars) are those stars from each group that have $\ali>2.65$ --- i.e., they have lithium abundances near or above the BBN \ali amount \citep[$\ali=2.75\pm0.02$;][]{Pitrou2018}. These stars are of interest for the accreted samples (GSE and the retrograde halo) as they could represent possible lithium enrichment in their formation environments. The solid black line is a thin disk evolutionary model from \citet{Cescutti2019} and the dashed black line is an evolutionary model for GSE from \citet{Cescutti2020}. \label{fig:possible_enrichment}}
\end{figure*}
Using the \teff selection from Section~\ref{sec:lithium} and the dynamical selections from Section~\ref{sec:ges_selection}, we now compare the properties of the Spite Plateau for sub-populations of the Milky Way. As discussed in Section \ref{sec:intro}, the aim of this work is to explore the lithium abundances of GSE stars and compare them to the rest of the Milky Way halo (a mixture of accreted and \textit{in situ} stars) and the Milky Way disk stars (likely \textit{in situ} formation), to see if there is any difference that could be the result of formation environment.

Figure~\ref{fig:possible_enrichment} compares the four populations defined in Section~\ref{sec:ges_selection}: GSE (first column), the retrograde halo (second column), the prograde halo (third column), and the disk (fourth column). The top row shows the \ali distribution with \feh, with other rows showing for context the \feh-\alphafe, \jphi-$\sqrt{\jr}$, and \teff-\logg distributions of the stars. For each population we only highlight the stars with $\teff>5850$~K (Section~\ref{sec:lithium}); additionally for the right-most column, the panels showing the disk stars, only those stars with $\feh<-0.6$ are shown for visual clarity (this metallicity was chosen as it is the observed metallicity ceiling of the other three samples). The purple square symbols further highlight stars within each sample that have $\ali>2.65$ --- stars that are near or above the BBN lithium abundance; these stars will be discussed in Section~\ref{sec:enrichment}.

As could already be clearly seen in Figure~\ref{fig:abundance_comparison}a and b, GALAH DR3 confirms the result from \citet{Molaro2020} and \citet{Cescutti2020} --- GSE shows broadly the same Spite Plateau lithium abundance as other stars in the Milky Way halo. Figure~\ref{fig:possible_enrichment} shows this result even more clearly with the sample limited to stars with $\teff>5850$~K: GSE stars, and the retrograde halo stars, are confined to a small region of \ali. The prograde halo and the disk sample show a larger range of \ali, but if we look at only the stars with $\feh<-1.3$, these stars are constant \ali with \feh.

To quantify the Spite Plateau \ali abundance, we consider only stars with $\feh<-1.3$, because stars above this metallicity in the prograde halo and disk samples begin to show a divergence from the Spite Plateau. This divergence is likely the signature of Galactic lithium evolution \citep{Bensby2018}, and not self-depletion or -enrichment of lithium in these stars. A bootstrap method was used to estimate the mean and standard deviation for each population. For each sample of stars, we re-sampled the \ali values with replacement 1000 times. For each re-sampled set of values we found the mean and the standard deviation, and then found the mean and the standard deviation of these 1000 values. On the top row of Figure~\ref{fig:possible_enrichment} the coloured horizontal line is the mean of the 1000 mean values, and the shaded region indicates the average of the 1000 standard deviations.

For all 251 GALAH dwarf stars with $\teff>5850$~K, $\feh<-1.3$, and measured lithium abundance, we find the Spite Plateau has a mean of $\ali=2.35\pm0.01$ and a spread of $\sigma_{\ali}=0.12\pm0.01$. For the four sub-sets of stars:
\begin{itemize}
	\item 37 GSE stars: $\ali=2.37\pm0.02$, $\sigma_{\ali}=0.12\pm0.02$;
	\item 45 retrograde halo stars: $\ali=2.37\pm0.02$, $\sigma_{\ali}=0.15\pm0.03$;
	\item 34 prograde halo stars: $\ali=2.35\pm0.02$, $\sigma_{\ali}=0.10\pm0.01$;
	\item 40 disk stars: $\ali=2.33\pm0.02$, $\sigma_{\ali}=0.14\pm0.02$.
\end{itemize}
The metallicity range for which GALAH provides reliable metallicities is $\feh>\sim-3$, and we confirm \cite[as has previously been seen, e.g.,][]{Rebolo1988,Melendez2010} that the Spite Plateau is basically flat with metallicity for stars within this metallicity regime. The four sub-samples of stars show essentially identical mean \ali abundances.

To summarize, we do not find any evidence to support the hypothesis that the formation environment affects the Spite Plateau. This result is not overly surprising. While, it was previously difficult to identify stars in the Milky Way that truly formed in another galaxy, there was a consensus that a large fraction of the halo was accreted during the hierarchal mergers. So the prior literature on the Spite Plateau \citep[e.g.,][]{Pinsonneault1992,Pinsonneault1999,Ryan1999,Melendez2004,Bonifacio2007} will have certainly contained some mixture of \textit{in situ} and \textit{ex situ} stars. If there was some large difference in the formation lithium abundance of stars that was driven by their host galaxy, then this should have been obvious as a large scatter in the Spite Plateau --- something that is not seen at the metallicities considered in this work.

\section{Metal-poor stars above the Spite Plateau} \label{sec:enrichment}
In Figure~\ref{fig:possible_enrichment} we have highlighted with purple squares those metal-poor stars in each population that show \ali near or above the BBN lithium abundance. Most of these stars are found in the disk sample, and all of the Li-rich disk dwarfs are found on the warm peninsula (Figure~\ref{fig:possible_enrichment}p). As discussed in \citet{Gao2020}, metal-poor stars with these properties could be stars that have truly retained the BBN abundance.

\begin{figure}
	\includegraphics[width=\columnwidth]{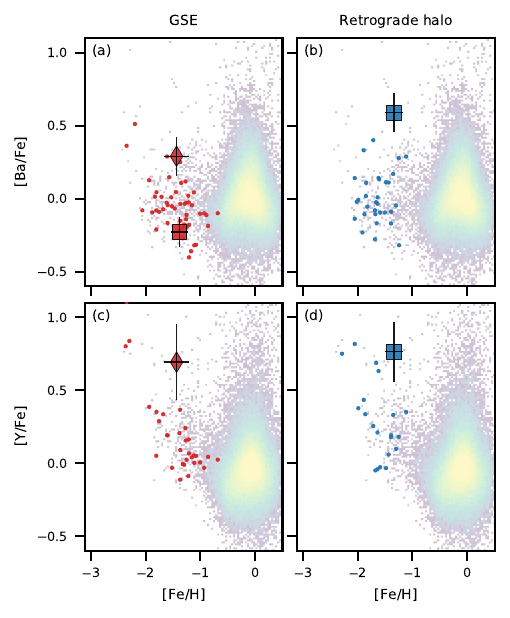}
    \caption{{The s-process abundances of the lithium-rich halo dwarf stars found in GALAH. The symbols are the same as Figure~\ref{fig:possible_enrichment}. The lack of stars in the lower-left quadrant of the panels is related to detection limitations of HERMES spectra. There does not appear to be any correlation between lithium and s-process enrichment. Two of the stars with high Li abundances have high s-process abundances, but two have either normal-to-low or no measurable s-process abundances.} 
     \label{fig:s_process_lithium_stars}}
\end{figure}

For the halo and GSE samples, there are four metal-poor stars that sit well above the Spite Plateau. Unlike the hot ($\teff > 6500$~K) stars of the disk mentioned above, these stars are cooler and so should sit on the Spite Plateau. Such Li-rich stars have been observed in globular clusters \citep{Koch2011,Monaco2012} and in the field \citep{Bonifacio1997,Asplund2006}. Within the framework for the Spite Plateau proposed by \citet{Fu2015} --- lithium depletion by the stars, followed by accretion from the interstellar medium --- such lithium-rich stars could also be the result of an increase of the accretion process.

They could be the result of mass transfer from an AGB companion. {The GALAH \texttt{flag\_sp} includes bits to signify stars that could be binaries --- either because they sit on the equal mass binary sequence, or are found co-located with other binaries in the t-SNE projections. However, all the stars considered in this work have $\texttt{flag\_sp}==0$, so were not flagged as possible binaries. One of the stars does have a large \vbroad, indicating it has broad spectral lines --- this could be a possible signature of binarity. If these stars were the product of mass transfer from an AGB companion, then this might manifest in the s-process abundances, which are shown for these stars in Figure~\ref{fig:s_process_lithium_stars}. Interpretation of neutron-capture element abundances in GALAH at low metallicity should be treated with caution as there is a strong metallicity-dependent detection limit. For the very lithium-rich retrograde halo star (blue square), it is clearly Ba- and Y-enhanced --- though there are also several ``lithium-normal'' stars of similar s-process abundances. Conversely, the most lithium-rich GSE star has low Ba and undetected Y.} 

Alternatively, it could represent lithium enrichment of the interstellar medium from which the stars formed. Models and observations of the Milky Way have its overall \ali increasing at around $\feh\sim-0.8$ \citep{Bensby2018,Cescutti2019}. This increase is driven by carbon-oxygen classical novae which produce \nuclide{7}{Be} that then decays to \nuclide{7}{Li} \citep{Starrfield2020,Grisoni2019}. \citet{Cescutti2020} and \citet{Molaro2020} both investigated evidence for lithium enrichment in GSE, with their models predicting the enrichment of lithium would occur at $\feh\sim-1.2$, a lower metallicity than in the Milky Way. Figure~\ref{fig:possible_enrichment} shows an exemplar Galactic lithium evolution model from \citet{Cescutti2019} as the solid black curve, and Figure~\ref{fig:possible_enrichment}a also has a lithium enrichment model for GSE from \citet{Cescutti2020} as the dashed curve. These novae have similar progenitors (WDs in binaries) as Type Ia SNe, \citet{Cescutti2020} assume a similar time delay of $\sim1$~Gyr before the onset of novae as typically assumed for Type Ia SNe. Thus an upturn in lithium abundance should occur at around the same time (i.e., \feh) as the downturn in \alphafe. We clearly see this downturn in \alphafe (Figure \ref{fig:abundance_comparison}c), but there is very much a lack of lithium-rich GES stars (Figure \ref{fig:possible_enrichment}a). There are only two GSE stars with enhanced lithium at lower metallicities, but they both at $\feh<-1$, {lower than the metallicity predicted by \citet{Cescutti2020} for the Li-enrichment}. Both can be seen to have orbital actions consistent with GSE, and one definitely has a low $\upalpha$ abundance, so there is nothing to rule out their GSE membership. GES is not the only accretion contributor to the halo of the Milky Way. So, there would be a variety of formation galaxy masses that have contributed. We might therefore expect to find many metal-poor stars above the Spite Plateau, but there are just four in our sample of 251 stars.

\section{Summary}\label{sec:discussion}

One of the main aims of Galactic archaeology, and in particular the GALAH survey, is to link the observations of the Milky Way to those of other galaxies --- the concept of ``near-field cosmology'' \citep{Freeman2002}. Within the Milky Way we have the ability to measure the properties and abundances of millions of individual stars to great precision, something that is only possible for some of the most luminous stars in nearby galaxies.

The \gaia revolution has shown the Milky Way contains a plethora of spatial and kinematic substructure \citep[see the review by][]{Helmi2020}, confirming the model of hierarchical galaxy formation. One of the dominant features is GSE, postulated to be the accretion remnant of a dwarf galaxy with a mass similar to the present-day Large Magellanic Cloud. It provides a large sample of identifiable \textit{ex-situ} formed stars, readily observable with moderate-sized telescopes. Every Galactic stellar survey with a simple selection that includes halo stars will have a large sample of GSE stars.

In this work we have used results from the Third Data Release of the GALAH survey to investigate the lithium abundance of $\sim100$ GSE stars. In particular we are interested in whether the formation environment of stars could be part of the solution to the cosmological lithium problem --- the observed discrepancy between the primordial amount of lithium predicted to have been formed by Big Bang Nucleosynthesis and the amount of lithium observed in warm, metal-poor stars \citep[][]{Fields2011}.

The large dataset of GALAH makes it possible to carefully consider the effects of stellar parameters on the observed lithium abundance of stars (Section \ref{sec:lithium}). The lithium abundance of a star cannot be considered in isolation from the observed stellar parameters, as standard models of stellar evolution do not successfully predict the observed trends between the lithium abundance of a star and its \teff, \logg, and \feh. As shown in Figure~\ref{fig:lithium_selection}, there are features such as the lithium dip region and the warm peninsula, which were studied in detail with the same data by \citet{Gao2020}. For the purposes of studying the Spite Plateau here we require stars to have $\teff>5850$~K, because this is the region of parameter space that contains stars least affected by main-sequence lithium depletion.

To identify stars from the GSE we used orbital angular momenta and integrals of motion. Figure~\ref{fig:initial_ges_selection_dynamics} shows that GSE stars are found to have large \jr and $\jphi\approx0~\jphiunits$. Kinematic substructure cannot be identified as accreted solely from its position in dynamical space. \citet{Jean-Baptiste2017} showed with high-resolution, dissipation-less N-body simulations that in various dynamical parameter spaces there is considerable overlap between the \textit{in-situ} and the accreted populations. Figures~\ref{fig:Feuillet_reproduction} and \ref{fig:abundance_comparison} show that our GSE selection does appear to be relatively free from contamination by obvious \textit{in-situ} stars --- the metallicity CDF are very similar for the prograde and retrograde populations of stars with $\jr>30~\jrunits$; and our GSE stars with $\feh>-1$ show low $\upalpha$-element abundances as expected for stars that formed in a low galactic mass system.

The main conclusion of this work is that the formation environment of a warm, metal-poor halo star does not play a role in its main sequence lithium abundance. There is no obvious difference in the lithium abundance, or its scatter, for metal-poor ($-3<\feh<-1.3$) stars from the GSE compared to stars from the halo or disk. In particular, stars from the retrograde halo --- which represent a stellar population that is likely to be either accreted or formed in the ancient proto-Galaxy --- are indistinguishable from the GSE stars. This fits with the framework that sees the Spite Plateau as the consequence of a lithium depletion by stars themselves.

This work used the motivation of the model proposed by \citet{Piau2006} that the Spite Plateau is simply the result of the first generation of stars of a galaxy efficiently depleting lithium. We extended this to a more general idea that, like $\upalpha$-element abundances, different galaxies would leave a different imprint on the lithium abundances. It should be noted that this proposed scenario has been rejected \citep[e.g.,][]{Prantzos2010,Prantzos2012} on the basis as it requires processing over two-thirds of all baryonic matter in very short-lived stars prior to any of the present-day halo stars formed, including at $\feh<-3$. Such massive star formation would produce large amounts of metals when exploding as super- and hyper-novae and thus raise the metallicities to well above those of the halo stars.

The other notable result from that there are four stars in the halo or GSE with lithium abundances near or above the BBN value. A handful of such stars have been seen before \citep{Bonifacio1997,Asplund2006,Koch2011,Monaco2012}. It is possible that they are the consequence of post-formation accretion, from either a companion, or the interstellar medium from which they formed. 

\section*{Acknowledgements}

The GALAH survey is based on observations made at the Anglo-Australian  Telescope, under programmes A/2013B/13, A/2014A/25, A/2015A/19, A/2017A/18. We acknowledge the traditional owners of the land on which the AAT stands, the Gamilaraay people, and pay our respects to elders past, present and emerging. This paper includes data that have been provided by AAO Data Central (\url{datacentral.org.au}).

The following software and programming languages made this research possible: \textsc{python} (v3.9.1); \textsc{astropy} \citep[v4.2;][]{TheAstropyCollaboration2018a}, a community-developed core Python package for astronomy; \textsc{matplotlib} \citep[v3.3.3;][]{Hunter2007,Caswell2020}; \textsc{scipy} \citep[v1.6.0;][]{SciPy1.0Contributors2020}; and \textsc{h5py} (v3.1.0).

This work has made use of data from the European Space Agency (ESA) mission {\it Gaia} (\url{https://www.cosmos.esa.int/gaia}), processed by the {\it Gaia} Data Processing and Analysis Consortium (DPAC, \url{https://www.cosmos.esa.int/web/gaia/dpac/consortium}). Funding for the DPAC has been provided by national institutions, in particular the institutions participating in the {\it Gaia} Multilateral Agreement.

Parts of this research were conducted by the Australian Research Council Centre of Excellence for All Sky Astrophysics in 3 Dimensions (ASTRO 3D), through project number CE170100013. JDS, SLM and DZ acknowledge the support of the Australian Research Council through Discovery Project grant DP180101791. SLM and JDS are supported by the UNSW Scientia Fellowship program. KL acknowledges funds from the European Research Council (ERC) under the European Union’s Horizon 2020 research and innovation programme (Grant agreement No. 852977).

\section*{Data availability}
The data underlying this article are available in the AAO Data Central. 

The GALAH DR3 catalogue, several value-added catalogues, and all HERMES spectra of the sources are available for download via the Data Central service at \url{datacentral.org.au}. The accompanying documentation can be found at \url{docs.datacentral.org.au/galah}, and a full description of the data release is given in \citet{Buder2021}.







\bsp	
\label{lastpage}
\end{document}